\documentclass{elsart}

\usepackage{graphics}
\usepackage{graphicx}

\usepackage{amssymb}

\newcommand{\lsim}{\,{\buildrel < \over {_\sim}}\,}
\newcommand{\gsim}{\,{\buildrel > \over {_\sim}}\,}
\newcommand{\sqrtsNN}{\sqrt{s_{\scriptscriptstyle{{\rm NN}}}}}
\newcommand{\av}[1]{\left\langle #1 \right\rangle}

\newcommand{\mev}{\mathrm{MeV}}
\newcommand{\gev}{\mathrm{GeV}}
\newcommand{\tev}{\mathrm{TeV}}
\newcommand{\fm}{\mathrm{fm}}

\newcommand{\mum}{\mathrm{\mu m}}

\newcommand{\PbPb}{\mbox{Pb--Pb}}
\newcommand{\pPb}{\mbox{p--Pb}}
\newcommand{\NN}{\mbox{nucleon--nucleon}}

\newcommand{\RAA}{R_{\rm AA}}

\newcommand{\RDh}{R_{{\rm D}/h}}
\newcommand{\pt}{p_{\rm t}}
\renewcommand{\d}{{\rm d}}
\newcommand{\dEdx}{{\rm d}E/{\rm d}x}

\newcommand{\ccbar}{\mbox{$\mathrm {c\overline{c}}$}}
\newcommand{\bbbar}{\mbox{$\mathrm {b\overline{b}}$}}

\newcommand{\Dz}{\mbox{$\mathrm {D^0}$}}

\let\Otemize =\itemize
\let\Onumerate =\enumerate
\let\Oescription =\description
\def\Nospacing{\itemsep=0pt\topsep=0pt\partopsep=0pt\parskip=0pt\parsep=0pt}
\def\Topspac{\vspace{-0.5\baselineskip}}
\def\Botspac{\vspace{-0.2\baselineskip}}

\begin{document}

\begin{frontmatter}



\hyphenation{author another created financial paper re-commend-ed Post-Script}


\title{Charm and beauty at the LHC}
\author{Andrea Dainese}
\address{INFN -- Laboratori Nazionali di Legnaro,\\ 
viale dell'Universit\`a 2, 35020, Legnaro (Padova), Italy}
\ead{andrea.dainese@lnl.infn.it}
       

\maketitle

\begin{abstract}
The Large Hadron Collider at CERN will open a new energy domain 
for heavy-ion physics. Besides ALICE, the dedicated heavy-ion 
experiment, also ATLAS and CMS are preparing rich physics programs
with nucleus--nucleus collisions.
Here we focus on open heavy-flavour and quarkonia studies, among the fields
that will most benefit from the high centre-of-mass energy at the LHC.
We discuss a few examples of physics issues that can be addressed
and we present a selection and comparison (where possible) 
of results on the expected capability of the three experiments. 
\end{abstract}

\begin{keyword}
Large Hadron Collider \sep heavy-ion collisions \sep heavy-flavour production
\PACS 25.75.-q \sep 14.65.Dw \sep 13.25.Ft
\end{keyword}

\end{frontmatter}

\section{Introduction}
\label{intro}

The nucleon--nucleon c.m.s. energy for \PbPb~collisions
at the LHC, $\sqrtsNN=5.5~\tev$, will exceed that available at RHIC by 
a factor about 30, opening up a new domain for the study of 
strongly-interacting matter in conditions of high energy 
density (QCD medium). 
High initial temperature is expected to characterize the system; 
the predicted values are on the order of $600~\mev$, 
twice as large as at RHIC and four times
larger than the critical temperature, $T_{\rm c}\approx 170~\mev$ 
at RHIC and LHC conditions, at which 
lattice QCD calculations predict the phase transition to a deconfined 
state of matter~\cite{karsch}.

At the LHC, 
hard-scattering processes should contribute significantly to the total 
cross section. The mechanism of energy loss 
due to medium-induced gluon radiation allows to use the energetic
partons produced in initial hard-scattering processes as probes 
to collect information on the opacity and density of the medium itself. 
The set of available probes will be extended both 
quantitatively and qualitatively. In fact, hard (light)quarks and 
gluons will be produced 
with high rates up to very large transverse momentum ($\pt$). 
Additionally, charm and beauty quarks, which, due to their masses, 
would show different attenuation patterns~\cite{carlosmagdalena} 
(section~\ref{pheno}), will become 
available for detailed measurements, since their production cross 
sections are expected to increase by factors 10 and 100, respectively,
from RHIC to the LHC. 
The entire spectrum of charmonia and bottomonia will be abundantly 
produced and a comparative measurement of their yields
is expected to provide insight on the 
properties, in particular on the temperature, of the deconfined medium
(section~\ref{pheno}).

\section{Heavy-quark and quarkonia phenomenology at the LHC}
\label{pheno}

\begin{table}[!b]
\caption{Expected $\rm Q\overline Q$ yields at the LHC, 
         from NLO pQCD~\cite{notehvq}. For \mbox{p--Pb}
         and \mbox{Pb--Pb}, 
         PDF shadowing is included and binary scaling is applied.}
\label{tab:xsec}
\begin{center}
\begin{tabular}{ccccc}
\hline
colliding system & $\sqrtsNN$ & centrality & $N^{\rm c\overline{c}}$/event  &  $N^{\rm b\overline{b}}$/event  \\
\hline
pp & $14~\tev$ & minimum bias  &  0.16 & 0.0072 \\
\pPb & $8.8~\tev$ & minimum bias & 0.78 & 0.029 \\
\PbPb & $5.5~\tev$ & central (0--5\% $\sigma^{\rm tot}$) & 115 & 4.6 \\
\hline
\end{tabular}
\end{center}
\end{table}

Heavy-quark pairs, $\rm Q\overline Q$, are produced in hard scattering
processes
with large momentum transfer, $Q^2\gsim (2\,m_{\rm Q})^2$. 
Hence, the 
production cross sections in \NN~(NN) collisions can be calculated 
in the framework of perturbative QCD (pQCD). 
The expected yields in pp collisions at $\sqrt{s}=14~\tev$
are reported in the first line of Table~\ref{tab:xsec}~\cite{notehvq},
as obtained at next-to-leading order 
using the MNR program~\cite{hvqmnr} with {\it reasonable} parameter values 
(quark masses and perturbative scales)~\cite{notehvq}. 
These yields have large uncertainties, of about a factor 2,
estimated by varying the parameters~\cite{notehvq}. 

For hard processes, in the absence of nuclear 
and medium effects, a nucleus--nucleus collision 
would behave as a superposition of independent NN collisions. 
The charm and beauty
yields would then scale from pp to AA 
proportionally to the number $N_{\rm coll}$ 
of inelastic binary NN collisions.
Binary scaling is, indeed, expected to break down due to initial-state 
effects, such as nuclear shadowing of the parton distribution functions 
(gluon recombination
in the high-density small-$x$ regime), 
and final-state effects, 
such as parton energy loss in the medium formed in AA collisions.

In Table~\ref{tab:xsec} we report the $\ccbar$ and $\bbbar$ yields 
in \pPb~and \PbPb~collisions, obtained including in the NLO pQCD
calculation the EKS98 parameterization~\cite{eks} of the PDFs nuclear
modification
and applying binary scaling~\cite{notehvq}. The charm (beauty) 
cross-section reduction induced by shadowing is about 35\% (20\%) in 
\mbox{Pb--Pb} and 15\% (10\%) in \mbox{p--Pb}. There is a significant 
uncertainty on the strength of shadowing in the small-$x$ region and 
some models predict much larger suppression than EKS98 
(see~\cite{yrpA} for a review). The comparison of ${\rm Q\overline Q}$ 
production in 
pp and \mbox{p--Pb} collisions 
is regarded as a sensitive tool
to probe nuclear PDFs. In particular,
the ratio of invariant-mass spectra of dileptons from heavy-quark decays 
for the two systems would measure 
the PDFs nuclear modification~\cite{yrpA}. 

  
Experiments at RHIC have shown that the nuclear modification factor of 
particles $\pt$ distributions, 
\mbox{$R_{\rm AA}(\pt,\eta)=
{1\over \av{N_{\rm coll}}} \cdot 
{\d^2 N_{\rm AA}/\d\pt\d\eta \over 
\d^2 N_{\rm pp}/\d\pt\d\eta}$},
is an effective observable 
for the study of the interaction of the hard partons 
with the medium produced in nucleus--nucleus collisions.
Heavy-quark medium-induced quenching is one of the most captivating 
topics to be 
addressed in \PbPb~collisions at the LHC. Due to the 
QCD nature of parton energy loss, quarks are predicted to lose less
energy than gluons (that have a higher colour charge) and, in addition, 
the `dead-cone effect' is expected to reduce the energy loss of massive 
quarks~\cite{dk,asw}. Therefore, one should observe a pattern 
of gradually decreasing $\RAA$ suppression when going from gluon-originated
light-flavour hadrons ($h^\pm$ or $\pi^0$) 
to D and to B mesons~\cite{carlosmagdalena,adsw}: 
$\RAA^h\lsim\RAA^{\rm D}\lsim\RAA^{\rm B}$. The enhancement with respect to 
unity of {\it heavy-to-light $\RAA$ ratios}
have been suggested~\cite{adsw} as well-suited observables to test the 
colour-charge ($R_{{\rm D}/h}=\RAA^{\rm D}/\RAA^h$) and mass 
($R_{{\rm B}/h}=\RAA^{\rm B}/\RAA^h$) dependence of parton energy loss.\\
The azimuthal anisotropy of particle production in non-central events 
is regarded as 
a powerful tool to study the early stage of the 
collision.
At the LHC, the large cross section for heavy-quark production will 
allow the direct measurement of the charm and beauty mesons azimuthal 
anisotropy coefficient $v_2$ up 
to large transverse momenta.
Depending on the considered $\pt$ range, the measurement of the 
D and B mesons $v_2$ probes: 
(a) the degree of thermalisation of c and b quarks in the expanding 
medium, at low and intermediate momenta ($\lsim 10~\gev/c$), 
where elliptic flow would be induced by collective pressure effects;
(b) the in-medium path-length dependence of heavy-quark energy loss in the 
almond-shaped partonic system, at higher momenta ($\gsim 10~\gev/c$).

The measurement of D and B meson production cross sections will 
also serve as a baseline for the study of medium effects on quarkonia.
Two of the most interesting items in the quarkonia sector at the LHC 
will be: (a) understanding the 
interplay between suppression and regeneration for 
J/$\psi$ production in a medium containing on the order of 100 $\ccbar$ pairs; 
(b) measuring for the first time 
medium effects on the bottomonia resonances, expected to be available
with sufficient yields at the LHC. On this point, the predicted suppression 
pattern as a function of the plasma temperature is particularly
interesting (see~\cite{satz} and references therein): 
the $\Upsilon$ would melt at 
about 2.5~$T_{\rm c}\approx 420~\mev$,
a temperature that would be reached only at the LHC, while the 
$\Upsilon^\prime$ would melt at the same temperature as the J/$\psi$,
about 1.2~$T_{\rm c}$. It will thus be important for the experiments to be able 
to measure also the $\Upsilon^\prime$, because, at variance with the 
J/$\psi$, it is expected to have a small regeneration probability and it would 
be very useful to disentangle J/$\psi$ suppression and regeneration.
We refer to~\cite{satz} for a detailed review on quarkonia phenomenology.
Finally, we note that, in order to study the medium effects on charmonia, 
it will be mandatory to measure the fraction of secondary charmonia from B 
decays, expected to be about 20\% for J/$\psi$ and 40\% for $\psi^\prime$,
in absence of medium-induced effects.

\section{ALICE, ATLAS and CMS as heavy-flavour detectors}
\label{exp}

Three experiments will participate in the LHC heavy-ion program:
ALICE, the dedicated heavy-ion experiment~\cite{alicePPR,alicePPR2}; 
CMS, with a
strong heavy-ion program~\cite{cms,roland}; ATLAS, which only
recently expressed interest in participating~\cite{atlas,takai}. The three 
detectors have different features and design requirements, but all of them 
are expected to have excellent capabilities for heavy-flavour measurements.
Their complementarity will provide
a very broad coverage in terms of phase-space, decay channels and observables.

\begin{figure}[!t]
\begin{center}
\includegraphics[width=.45\textwidth]{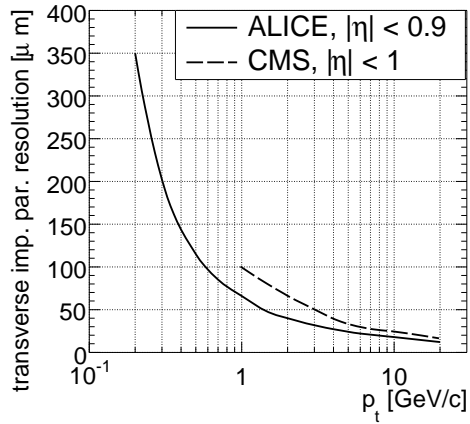}
\caption{Transverse track impact parameter
         resolutions for the ALICE~\cite{alicePPR2} 
         and CMS~\cite{roland} detectors in \mbox{Pb--Pb}
         collisions.}
\label{fig:d0}
\end{center}
\end{figure}

Experimentally, the two key elements for a rich heavy-flavour program are:
tracking/vertexing and particle identification.\\
Open charm and beauty mesons have typical life-times of few hundred microns
($c\tau$ values are about $125$--$300~\mum$ for D's and $500~\mum$ for B's) 
and the most direct detection strategy is the identification of single
tracks or vertices that are displaced from the interaction vertex.
The detector capability to perform this task is
determined by the impact parameter\footnote{We define as impact 
parameter the distance of closest approach to the interaction vertex 
of the track projection in the plane transverse to the beam direction.}
($d_0$) resolution. All experiments will be equipped with 
high position-resolution silicon-detector layers, including pixels, for
precise tracking and impact parameter measurement also in the 
high-multiplicity environment of central \PbPb~collisions.
Tracking is done in the central pseudorapidity region:
$|\eta|<0.9$ for ALICE and $|\eta|\lsim 3$ for CMS and ATLAS.  
In Fig.~\ref{fig:d0} we show the $d_0$ resolution of ALICE and 
CMS. The $d_0$ resolutions are 
quite similar and better than $50~\mum$ for $\pt\gsim 1.5$--$3~\gev/c$.
One of the main differences 
between the three experiments is given by the magnetic 
field value, that determines the $\pt$ resolution at high $\pt$ on one
hand, and the low-$\pt$ reach on the other.
ALICE (0.5~T) has a very low $\pt$ cutoff of $0.2~\gev/c$ and a $\pt$ 
resolution of about 5\% at $100~\gev/c$, while CMS (4~T) and ATLAS (2~T) have 
a higher cutoff of about $1~\gev/c$
and a resolution of 2\% at $100~\gev/c$.

Both lepton and hadron identification are important for heavy-flavour 
detection. D and B mesons have relatively-large branching ratios (BR) in the 
semi-leptonic channels, $\simeq 10\%$ to electrons and $\simeq 10\%$ to muons,
and inclusive cross-section measurements can be performed via single leptons 
or dileptons. 
Moreover, quarkonia production is measured using dimuon or dielectron
invariant-mass analyses.
ALICE can identify electrons with $\pt>1~\gev/c$ and 
$|\eta|<0.9$, via transition radiation and $\dEdx$ measurements, and muons 
in the forward region, $-4<\eta<-2.5$, which allows a very low $\pt$ cutoff 
of $1~\gev/c$. CMS and ATLAS have 
a broad pseudorapidity coverage for muons, $|\eta|<2.4$ and $|\eta|<2.7$,
respectively, but they have a higher momentum cutoff, 
$p\gsim 4~\gev/c$ (i.e. $\pt\gsim 4~\gev/c$ at $\eta=0$). 
Both CMS and ATLAS have high-resolution 
electromagnetic calorimeters that can be used to identify 
electrons, although performance studies for heavy-ion collisions have not
been carried out yet. Semi-leptonic inclusive measurements do not
provide direct information on the D(B)-meson $\pt$ distribution, especially
at low $\pt$, because of the weak correlation between the lepton and meson
momenta. Therefore, for charm in particular, the reconstruction of exclusive
(hadronic) decays is preferable. In this case, 
in a high-multiplicity environment, 
hadron identification allows a more effective rejection
of the large combinatorial background in the low-$\pt$ region.
ALICE disposes of $\rm \pi/K/p$ separation via $\dEdx$ and time-of-flight 
measurement for $p<3$--$4~\gev/c$ and $|\eta|<0.9$.

\section{Open charm and beauty capabilities}
\label{open}

In this section we present the expected capability of ALICE for 
the measurement of D and B meson production. Detailed studies 
for ATLAS and CMS have not been performed yet. However, given the
features that we have just discussed, 
these experiments can be expected to have similar performance as
ALICE, though with a more limited low-$\pt$ reach, because of 
the lack of hadron identification and of the $\approx 4~\gev/c$ 
momentum threshold for muons.

\begin{figure}[!t]
  \begin{center}
    \includegraphics[width=.45\textwidth]{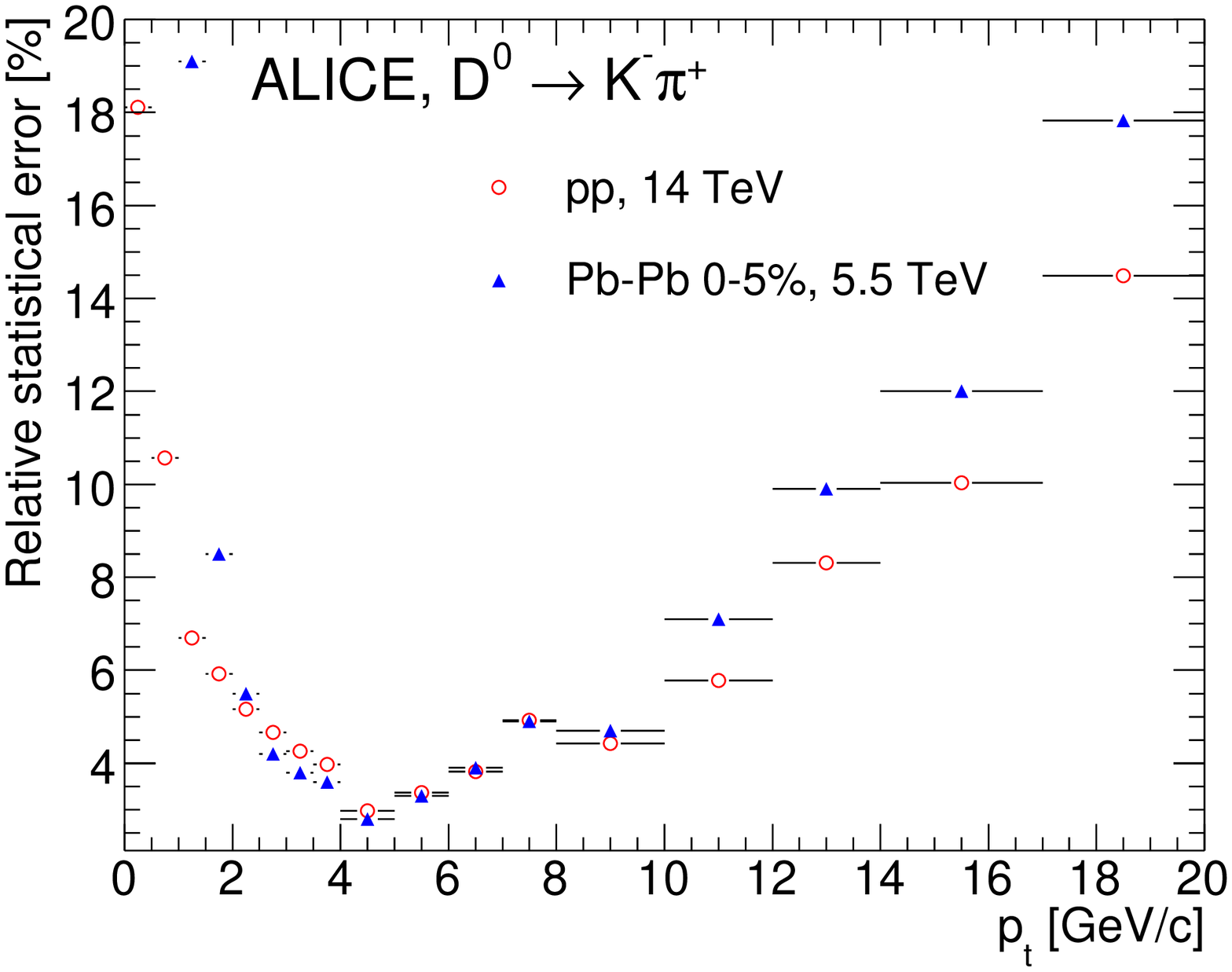}
    \includegraphics[width=.45\textwidth]{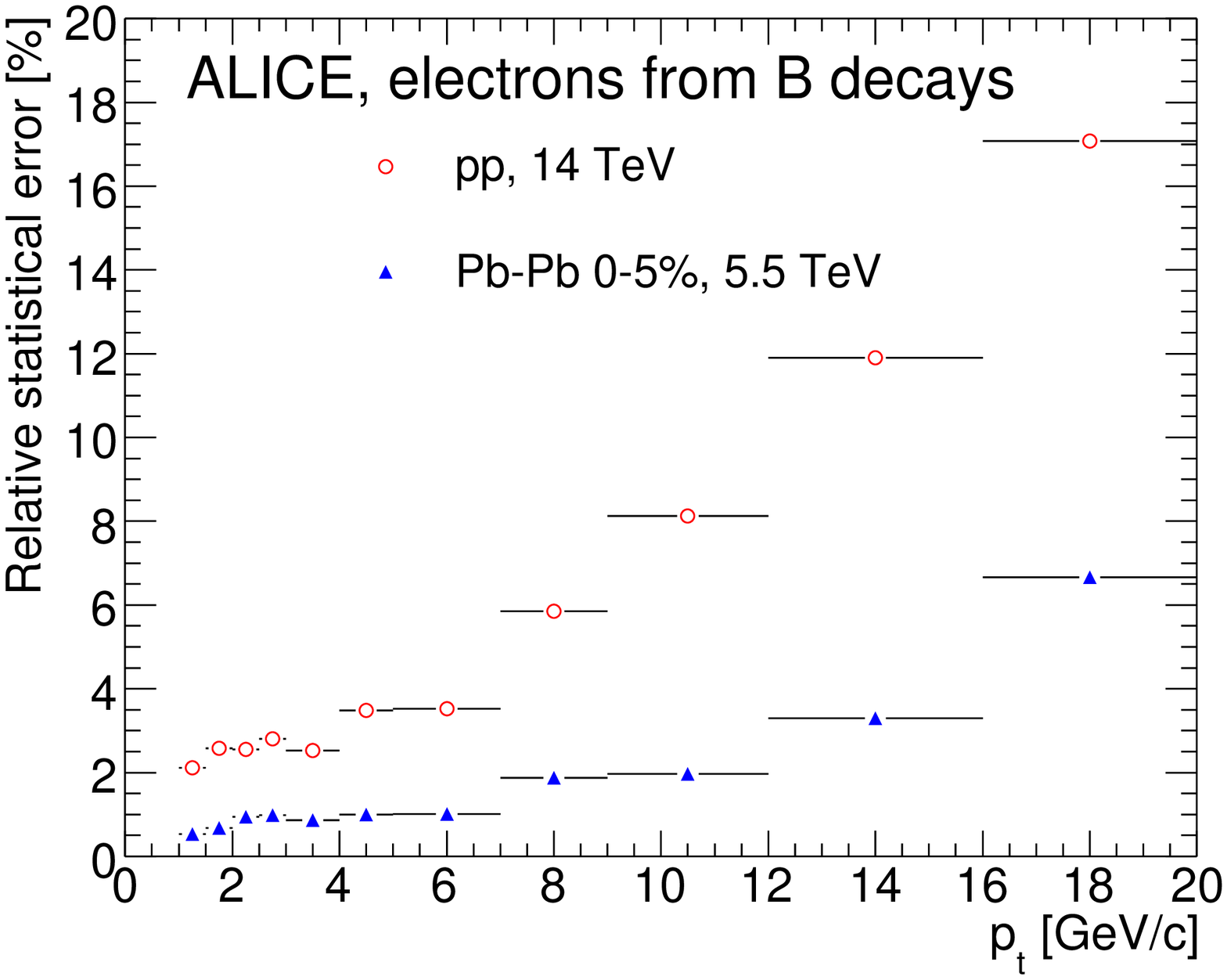}
    \caption{Expected relative statistical errors for the measurement 
             in ALICE
             of the production cross sections of ${\rm D^0}$ (left) and 
             of B-decay electrons (right), 
             in 0--5\% central Pb--Pb collisions 
             and in pp collisions.} 
    \label{fig:D0Bestat}
  \end{center}
\end{figure}

{\it Exclusive charm reconstruction in ALICE}.
One of the most promising channels for open charm detection is the 
$\rm D^0 \to K^-\pi^+$ decay (and charge conjugate) that 
has a BR of $3.8\%$.
The main feature of this decay topology is the presence of two tracks with 
impact parameters $d_0\sim 100~\mum$. The detection strategy
to cope with the large combinatorial background from the underlying event 
is based on the selection of displaced-vertex topologies, i.e. two tracks with 
large impact parameters and good alignment between the $\rm D^0$ momentum 
and flight-line, and on invariant-mass analysis to extract the signal 
yield~\cite{alicePPR2}.
As shown in Fig.~\ref{fig:D0Bestat} (left),
the accessible $\pt$ range is $1$--$20~\gev/c$ for Pb--Pb and 
$0.5$--$20~\gev/c$ for pp, 
with statistical errors better than 15--20\% at high $\pt$. 
The systematic errors 
(acceptance and efficiency corrections, 
centrality selection for Pb--Pb) are expected to be smaller than 20\%.

{\it Beauty via single electrons in ALICE.}
The main sources of background electrons are: decays of D mesons; 
$\pi^0$ Dalitz decays 
and decays of light vector mesons (e.g.\,$\rho$ and $\omega$);
conversions of photons in the beam pipe or in the inner detector 
layer and pions misidentified as electrons. 
Given that electrons from beauty have average 
impact parameter $d_0\simeq 500~\mum$
and a hard momentum spectrum, it is possible to 
obtain a high-purity sample with a strategy that relies on:
electron identification with a combined $\dEdx$ and transition
radiation selection;
impact parameter cut to reject misidentified $\pi^\pm$ and $\rm e^{\pm}$
from Dalitz decays and $\gamma$ conversions, and to 
reduce the charm-decay component.
As an example, with $d_0>200~\mum$ and $\pt>2~\gev/c$, the expected statistics
of electrons from b decays is $8\times 10^4$ for $10^7$ central 
Pb--Pb events, allowing the measurement of electron-level 
$\pt$-dif\-fe\-ren\-tial 
cross section in the range $2<\pt<20~\gev/c$. 
In Fig.~\ref{fig:D0Bestat} (right) we show the expected statistical 
errors on the measurement of the cross section of electrons from b 
decays.

{\it Beauty via muons in ALICE.}
B production in \mbox{Pb--Pb} collisions 
can be measured also in the ALICE muon 
spectrometer ($-4<\eta<-2.5$) analyzing the single-muon $\pt$ 
distribution~\cite{alicePPR2}.
The main backgrounds to the `beauty muon' signal are $\pi^\pm$, 
$\rm K^\pm$ and charm decays. The cut $\pt>1.5~\gev/c$ is applied to all
reconstructed muons in order to increase the signal-to-background ratio.
Then, a fit technique allows to extract a $\pt$ distribution of muons 
from B decays.
Since only minimal cuts are applied, the statistical errors are 
expected to be smaller than 5\% up to muon $\pt\approx 30~\gev/c$.

\begin{figure}[!t]
  \begin{center}
    \includegraphics[width=0.49\textwidth]{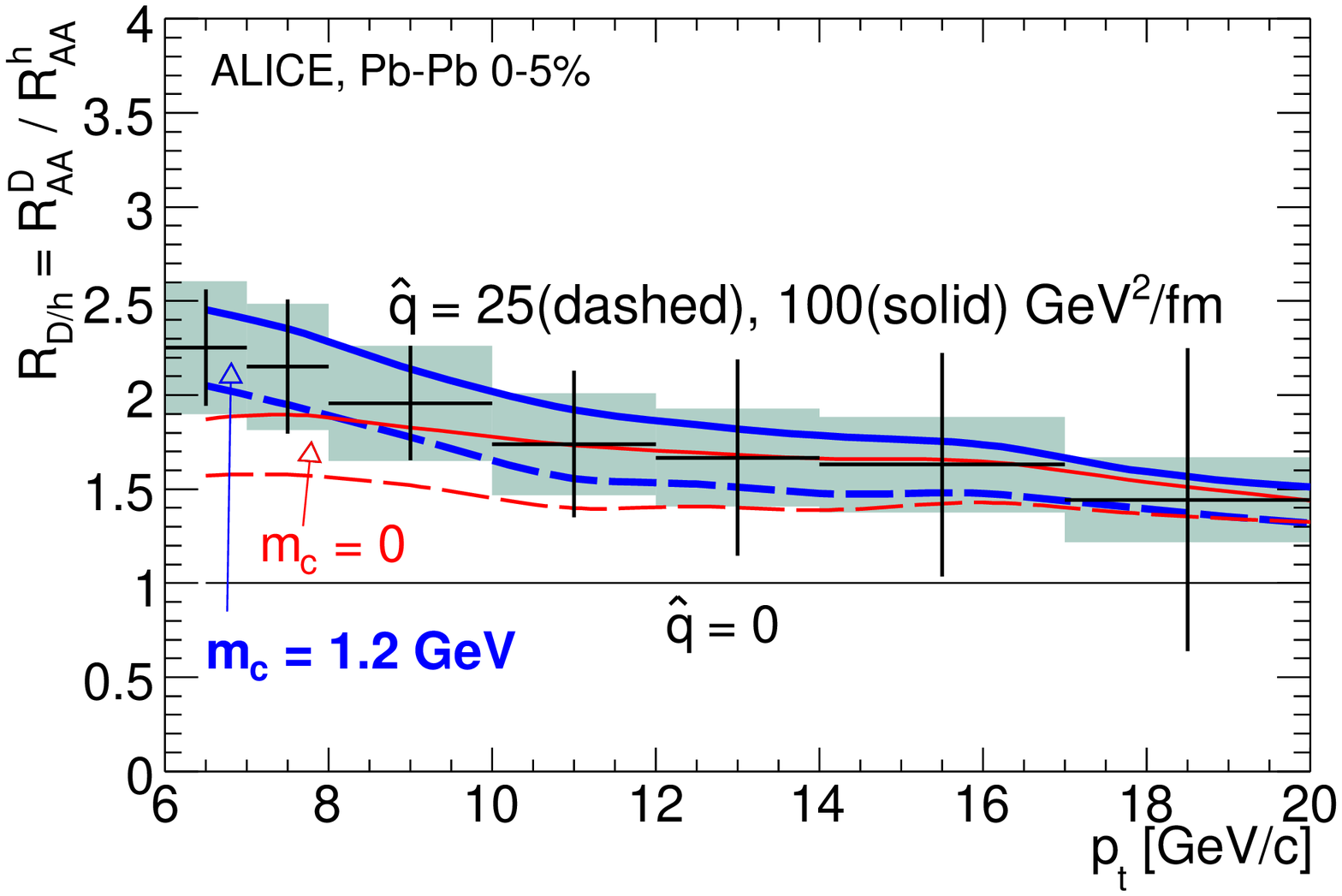}
    \includegraphics[width=0.49\textwidth]{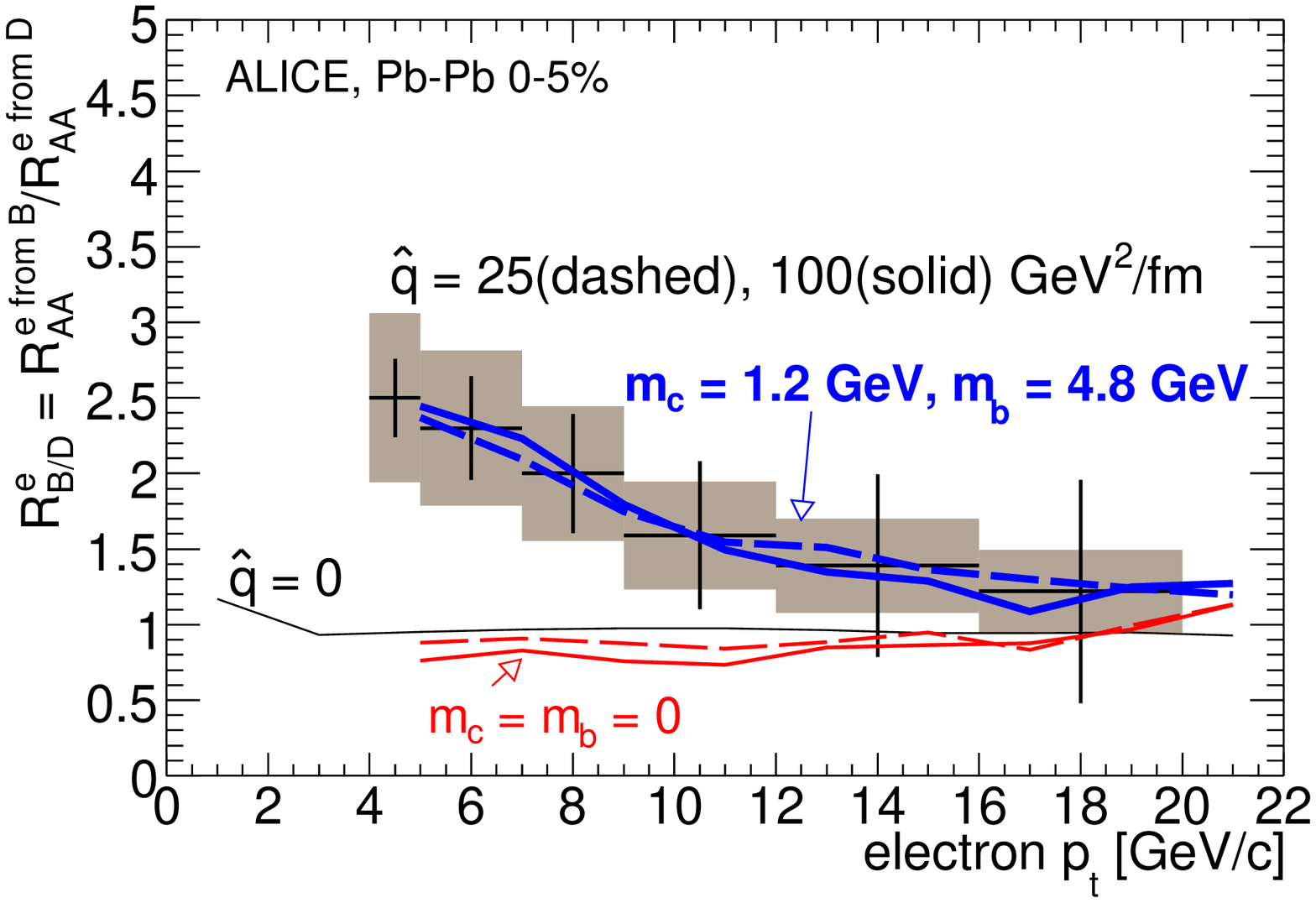}
    \caption{Ratio of the nuclear modification factors for $\Dz$ mesons 
             and for charged hadrons (left) and ratio of the nuclear
             modification factors for B-decay and for D-decay electrons
             (right).
             Errors corresponding to the centre of the prediction bands   
             for massive quarks are shown: bars = statistical, 
             shaded area = systematic.} 
    \label{fig:heavytolight}
  \end{center}
\end{figure}

{\it Heavy-to-light ratios in ALICE.}
ALICE investigated the possibility of using 
the described charm and beauty measurements 
to study the dependences of parton energy loss.
The expected experimental errors 
on these observables are compared to recent theoretical predictions from
parton energy loss~\cite{adsw}.
The sensitivity to the heavy-to-light ratios $\RDh=\RAA^{\rm D}/\RAA^h$ 
and $R_{\rm B/D}=\RAA^{\rm e~from~B}/\RAA^{\rm e~from~D}$ in the 
range $5<\pt<20~\gev/c$ is presented in Fig.~\ref{fig:heavytolight}
(the $\pt$ distribution of D-decay electrons will be calculated from 
the measured $\Dz$ $\pt$ distribution). 
Predictions with and without the effect of the
heavy-quark mass, for a medium transport coefficient in the range 
$25$--$100~\gev^2/\fm$, are shown.
For $5<\pt<10~\gev/c$, the measurement of the expected
enhancement of heavy-to-light ratios with respect to unity
appears to be feasible. 

\section{Quarkonia capabilities}
\label{quarkonia}

\begin{figure}[!t]
  \begin{center}
    \includegraphics[width=\textwidth]{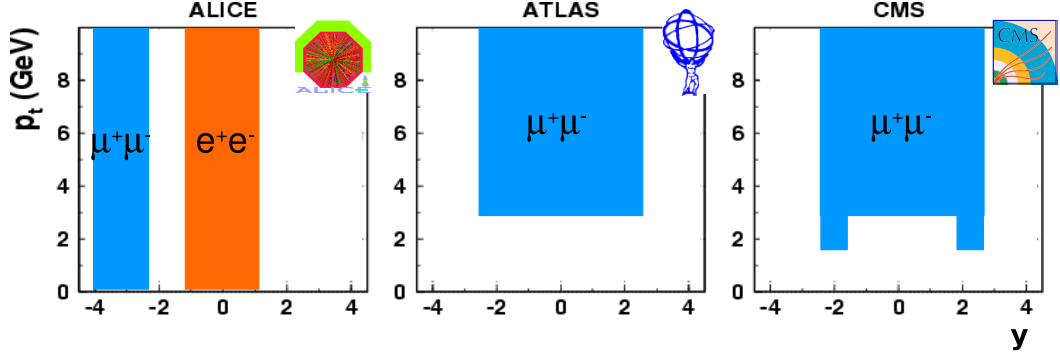}
    \caption{Schematic ($y$, $\pt$) quarkonia acceptances 
             for ALICE, ATLAS and CMS.} 
    \label{fig:accQQ}
  \end{center}
\end{figure}

\begin{figure}[!b]
  \begin{center}
    \includegraphics[width=\textwidth]{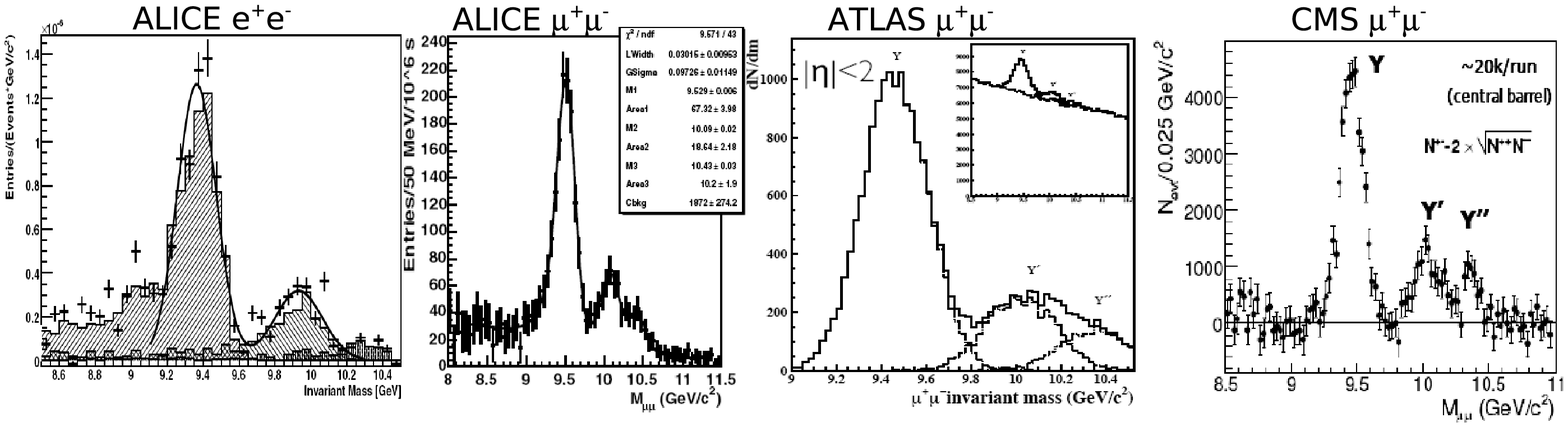}
    \caption{The $\Upsilon$ family in central \mbox{Pb--Pb} collisions, 
             as pictured by ALICE~\cite{alicePPR2}, ATLAS~\cite{takai} 
             and CMS~\cite{roland} in one month of data-taking.} 
    \label{fig:upsilon}
  \end{center}
\end{figure}

Figure~\ref{fig:accQQ} shows the schematic acceptances for charmonia and 
bottomonia in the ($y$,\,$\pt$) plane. ALICE can detect quarkonia 
in the dielectron channel at central rapidity ($|y|\lsim 1$) and in the
dimuon channel at forward rapidity ($-4<y<-2.5$). In both channels
the acceptance extends down to zero transverse momentum, since the 
minimum $\pt$ is $1~\gev/c$ for both electrons and muons. ATLAS and CMS
will use only dimuons and they have similar acceptances, covering 
$\pt\gsim 3~\gev/c$ and $|y|\lsim 2.5$. CMS and ATLAS studies indicate 
that, near the edges of the pseudorapidity window, there is some acceptance 
down to $\pt\approx 1.5~\gev/c$.
We emphasized the importance of separating the $\Upsilon$ and $\Upsilon^\prime$;
given that the mass difference between the two bottomonia is 
about $500~\mev$, a mass resolution of order $100~\mev$ at 
$M_{\ell^+\ell^-}\sim 10~\gev$, i.e. $\sigma_M/M\approx 1\%$, is required.
CMS and ALICE fulfill this requirement, while for ATLAS the 
$\Upsilon/\Upsilon^\prime$ separation will be more difficult.
We report in Table~\ref{tab:onia} a summary of the expected quarkonia
performances, including the mass resolution, the list of 
measurable states and the $\pt$ coverage for 
J/$\psi$ and $\Upsilon$.
For illustration, in Fig.~\ref{fig:upsilon} 
we show the simulated dilepton mass spectra in the 
$\Upsilon$ region after background subtraction~\cite{alicePPR2,takai,roland}. 
All three experiments will be able to measure the fraction of J/$\psi$
that feed-down from B decays, by studying the impact parameter distribution
of the dilepton pairs.

\begin{table}[!b]
\caption{Expected quarkonia performances in central \mbox{Pb--Pb}
         for ALICE, ATLAS and CMS.}
\label{tab:onia}
\begin{center}
\scriptsize
\begin{tabular}{clcccc}
\hline
&& ALICE ${\rm e^+e^-}$ & ALICE $\mu^+\mu^-$ & ATLAS $\mu^+\mu^-$ & CMS $\mu^+\mu^-$ \\
\hline
$\psi$ & $y$ acc. & $|y|<1 $ & $-4<y<-2.5$ & $|y|<2.5$ & $|y|<2$ \\
&Mass res. & 35~MeV &  65~MeV &  70~MeV & 37~MeV \\
&Meas. states & $\psi$ OK; $\psi^\prime$ ? & $\psi$ OK; $\psi^\prime$ ? & $\psi$ OK; $\psi^\prime$ ?& $\psi$ OK; $\psi^\prime$ ? \\
&$\pt$ range & 0--10~$\gev/c$ & 0--20~$\gev/c$ & not studied & few--20~$\gev/c$  \\
\hline
$\Upsilon$ & $y$ acc. & $|y|<1 $ & $-4<y<-2.5$ & $|y|<2.5$ & $|y|<0.8\,(2.4)$ \\
&Mass res. & 90~MeV &  90~MeV &  145~MeV & 54\,(85)~MeV \\
&Meas. states & $\Upsilon$ OK; $\Upsilon^\prime$ ? & $\Upsilon,\Upsilon^\prime$ OK; $\Upsilon^{\prime\prime}$ ? & $\Upsilon,\Upsilon^\prime$ OK; $\Upsilon^{\prime\prime}$ ? & $\Upsilon,\Upsilon^\prime$ OK; $\Upsilon^{\prime\prime}$ ? \\
&$\pt$ range & not studied & 0--8~$\gev/c$ & not studied & few--10~$\gev/c$  \\
\hline
\end{tabular}
\end{center}
\end{table}

\section{Summary}
\label{summary}

We have discussed how heavy quarks, abundantly produced at LHC energies, 
will allow to address several physics issues in heavy-ion collisions. 
In particular, they provide tools to probe the density, 
via parton energy loss and its predicted mass dependence,
and temperature,
via quarkonia successive dissociation patterns, 
of the high-density QCD medium formed in \mbox{Pb--Pb} collisions.
The excellent tracking, vertexing and particle identification performance 
of ALICE, ATLAS and CMS will allow to fully explore this rich phenomenology.

{\it Acknowledgment.} The author is grateful to F.~Antinori, P.~Crochet,
D.~d'Enter\-ria, G.~Martinez, 
A.~Morsch, G.~Roland, L.~Rosselet, 
J.~Schukraft, H~.Takai and B.~Wyslouch for valuable help
in the compilation of the experiments performance results.

\vspace{.4cm}


\begin{thebibliography}{99}

\bibitem{karsch}
  F.~Karsch, {\it these proceedings}.

\bibitem{carlosmagdalena}
  C.A.~Salgado, {\it these proceedings}; M.~Djordjevic, {\it these proceedings}.
\bibitem{notehvq}
  N.~Carrer and A.~Dainese, ALICE-INT-2003-019 (2003) [hep-ph/0311225]. 

\bibitem{hvqmnr} 
  M.~Mangano, P.~Nason and G.~Ridolfi, Nucl.~Phys.~{\bf B373} (1992) 295.

\bibitem{eks} 
  K.J.~Eskola {\it et al.}, Eur.~Phys.~J.~{\bf C9} (1999) 61.

\bibitem{yrpA}
  A.~Accardi {\it et al}, hep-ph/0308248, in CERN Yellow Report 2004-009.

\bibitem{dk}
  Yu.L.~Dokshitzer and D.E.~Kharzeev, Phys.~Lett.~{\bf B519} (2001) 199.

\bibitem{asw}
  N.~Armesto {\it et al.}, Phys.~Rev.~{\bf D69} 
  (2004) 114003.

\bibitem{adsw}
  N.~Armesto {\it et al.}, 
  Phys.~Rev.~{\bf D71} (2005) 054027.

\bibitem{satz}
  H.~Satz, {\it these proceedings}.

\bibitem{alicePPR}
  ALICE Coll., Physics Performance Report Vol.~I, CERN/LHCC 2003-049  
  and J.~Phys.~{\bf G30} (2004) 1517.

\bibitem{alicePPR2}
  ALICE Coll., Physics Performance Report Vol.~II, CERN/LHCC 2005-030  
  and J.~Phys.~{\bf G32} (2006) 1295.

\bibitem{cms}
  CMS Coll., Physics TDR Vol.~II, CERN/LHCC~2006-021.

\bibitem{roland}
  G.~Roland, {\it these proceedings}.
 
\bibitem{atlas}
  ATLAS Coll., Heavy-ion Letter of Intent, CERN/LHCC 2004-009 (2004).

\bibitem{takai}
  H.~Takai, {\it these proceedings}.



\end{thebibliography}
\end{document}